\documentclass[a4paper,superscriptaddress,twocolumn,prb,aps,showpacs,floatfix,citeautoscript]{revtex4}
\makeatletter
\def\@dotsep{4.5}
\makeatother

\usepackage{graphicx}
\usepackage{bm}
\usepackage{amsmath,amsbsy,amsfonts,mathrsfs}

\newcommand {\ord} [1] {^{\left(#1\right)}}
\newcommand {\der} {\left(\vec{r}\right)}
\newcommand {\dert} {\left(\vec{r}, t\right)}

\newcommand {\w} [1] {\omega_{#1}}
\newcommand {\xc} {_{\rm xc}}
\renewcommand {\vec} [1] {{\bm #1}}
\newcommand {\bpar} {\beta_{\parallel}}

\newcommand{\D}{{\rm d}}

\newcommand{\mint}[1]{\int\! \D^{3} #1 \, }
\newcommand{\mdint}[2]{\mint{#1}\!\!\!\mint{#2}}

\newcommand {\sanse}{Departamento de F{\'\i}sica de Materiales,
  Facultad de Ciencias Qu\'{\i}micas, UPV/EHU, Centro Mixto
  CSIC-UPV/EHU and Donostia International Physics Center, San
  Sebasti\'an, Espa\~na.}

\newcommand {\lsi}{Laboratoire des Solides Irradi\'es,
  CNRS-CEA-\'Ecole Polytechnique, Palaiseau, France.}

\newcommand {\coimbra}{Centro de F\'{\i}sica Computacional, Departamento de F{\'\i}sica, Universidade de
  Coimbra, Coimbra, Portugal.}

\newcommand {\etsf}{European Theoretical Spectroscopy Facility.}

\begin{document}

\title{A time-dependent density functional theory scheme for efficient
  calculations of dynamic (hyper)polarizabilities}

\date{\today}

\author{Xavier Andrade}
\email{xavier@tddft.org}
\affiliation{\sanse}
\affiliation{\lsi}
\affiliation{\etsf}

\author{Silvana Botti}
\affiliation{\lsi}
\affiliation{\coimbra}
\affiliation{\etsf}

\author{Miguel A.\,L. Marques}
\affiliation{\coimbra}
\affiliation{\etsf}

\author{Angel Rubio}
\affiliation{\sanse}
\affiliation{\etsf}

\begin{abstract}
  We present an efficient perturbative method to obtain both static
  and dynamic polarizabilities and hyperpolarizabilities of complex
  electronic systems. This approach is based on the solution of a
  frequency dependent Sternheimer equation, within the formalism of
  time-dependent density functional theory, and allows the calculation
  of the response both in resonance and out of resonance.
  Furthermore, the excellent scaling with the number of atoms opens
  the way to the investigation of response properties of very large
  molecular systems. To demonstrate the capabilities of this method,
  we implemented it in a real-space (basis-set free) code, and applied
  it to benchmark molecules, namely CO, H$_2$O, and paranitroaniline
  (PNA). Our results are in agreement with experimental and previous
  theoretical studies, and fully validate our approach.
\end{abstract}

\maketitle

\section{Introduction}

The optical properties of a material are essentially determined by the response
of the electrons to an external field. If this applied field is small, the
induced dipole of the system can be expanded in powers of the
field\cite{butcher,bloembergen}. The first order coefficient is the so-called
electric polarizability \(\alpha\). This quantity describes, e.g., the
dielectric properties of the material and how light is absorbed and emitted. In
second order we obtain the first hyperpolarizability $\beta$, that is
responsible for the processes of second-harmonic generation, optical
rectification and Pockles effect\cite{butcher}.  Higher order terms can be
related to other electro-optical effects like third-harmonic generation, the
Kerr effect, etc.  Finally, polarizabilities are called static or dynamic if
the perturbing field is static or frequency dependent.

The importance of the linear term, $\alpha$, is well known in Physics and
Chemistry\cite{bonin}. On the other hand, non-linear (i.e. beyond first order) optical
effects have gained quite some interest lately due to their technological
applications in opto-electronic devices. Nonlinear optical materials can be used
to convert light to shorter (bluer) wavelengths, which can be focused to a
smaller spot size. Shorter wavelength light sources would hence yield higher
density optical recording media (such as DVDs and CDs).  Other applications
include tunable light sources, image recognition systems and adaptive optics.

Several methods have been used to calculate (hyper)polarizabilities of
finite systems\cite{langhoff,shelton94,bishop,review_baroni,cardoso}.
In the static case, the simplest approach is finite differences
\cite{gready77} that uses the definition of the polarizabilities as
derivatives of the dipole (or of the total energy) with respect to the
applied field.  Calculations are performed at various (small) field
strengths, and the required derivatives are evaluated numerically.
This method is simple and straightforward to implement. However, it
requires many total energy evaluations, and these need to have a very
high precision to obtain reasonable numerical derivatives. Moreover,
it is not possible to generalize this idea to the dynamic case.

Another widely used approach is perturbation theory, of which more
than one flavor exists. In the sum-over-states method
\cite{langhoff,orr71} the (hyper)polarizabilities are written as an
infinite sum over occupied and empty states, that involves the
ground-state eigenvalues and dipole matrix e\-le\-ments.  In a similar
vein, one can obtain the po\-la\-ri\-za\-bi\-li\-ties from the
corresponding response functions written in the product basis of
occupied and empty states or in terms of Green's
functions\cite{senatore8910}. Note that these techniques can be used
both for static and dynamic response.  Although widely used by the
community, these methods have several shortcomings.  First, results
are often difficult to converge with the number (and quality) of the
empty states. Second, the scaling with the number of atoms is quite
unfavorable, making hard the application to the study of large system,
like nanostructures or molecules with biological interest.

A different technique also used to obtain both static and dynamic
linear polarizabilities is the direct solution of the time-dependent
Schr\"odinger equation in real time \cite{Yabana96}. In this way only
the occupied subspace is needed and the scaling with system size is
excellent [${\cal O}(N^2)$, where $N$ is the number of atoms].
However, this method cannot be easily generalized to extract
hyperpolarizabilities.

A very interesting approach that essentially solves the problems
mentioned above is the Sternheimer equation\cite{sternheimer}.
Although a perturbative technique, it avoids the use of empty states,
and has a quite good scaling with the number of atoms. This method has
already been used for the calculation of many response
properties\cite{review_baroni} like atomic vibrations (phonons),
electron-phonon coupling, magnetic response, etc. In the domain of
optical response, this method has been mainly used for static
response, although a few first principles calculations for low
frequency (far from resonance) (hyper)polarizabilities have
appeared\cite{senatore87,karna90a,vangisbergen97,bertsch01}.
Recently, a reformulation of the Sternheimer equation in a super
operator formalism was presented\cite{baroni_turbo}. When combined
with a Lanczos solver, it allows to calculate very efficiently the
first order polarizability for the whole frequency spectrum. However
the generalization of this method to higher orders is not
straight-forward.

In this Article, we propose a modified version of the Sternheimer
equation that is able to cope with both static and dynamic response in
and out of resonance.  The solution of the first-order Sternheimer
equation gives us access to both $\alpha$ and $\beta$. Higher order
polarizabilities can be obtained from an hierarchy of Sternheimer
equations. Exchange and correlation effects are treated at the level
of density functional theory (DFT)\cite{hk} for static
po\-la\-ri\-za\-bi\-li\-ties and time-dependent DFT
(TDDFT)\cite{tddft} for the dynamic case. Compared to other
quantum-chemistry approaches, density functional methods have a
somewhat lower accuracy but are lighter numerically, allowing the
study of much larger systems. In the present work we focus on finite
systems but the method has also been applied to periodic systems for
the non-resonant case\cite{rubio96}. Note that in the field of DFT,
the Sternheimer equation is often referred to as density functional
perturbation theory \cite{review_baroni}.

The rest of this Article is organized as follows. In
Sect.~\ref{sec:theory} we present the derivation of the
frequency-dependent Sternheimer equation and show how to obtain the
linear polarizability and first hyperpolarizability from its solution.
In the following section we give some details concerning the
implementation of our method. In Sect.~\ref{sec:applications} we apply
this theory to several test molecules, comparing our results to other
calculations and experiments. Finally we present our conclusions and a
brief outlook.

\section{Theory}
\label{sec:theory}
\subsection{Linear Response}

Within TDDFT, the quantum state of an interacting electronic system is
described by the time-dependent Kohn-Sham equations (atomic units will
be used unless explicitly stated)
\begin{equation}
  \label{eq:tdks}
  i\frac{\partial}{\partial{t}}\psi_m(\vec r, t)=
  H_{\rm KS}\left(t\right)\psi_m(\vec r, t)
  \,.
\end{equation}
The Kohn-Sham Hamiltonian is written as
\begin{equation}
  H_{\rm KS} = -\frac{\nabla^2}{2} + v_{\rm ext}(\vec r, t) + 
  v_{\rm Hartree}(\vec r, t) + v_{\rm xc}(\vec r, t)
  \,,
\end{equation}
where the first term corresponds to the kinetic energy, and the following ones
represent the external potential, the Hartree potential that describes the
classical interaction between the electrons, and the exchange-correlation term
that accounts for all non-trivial parts of the electron-electron interaction.
Note that the Hartree and exchange-correlation terms are time-dependent as they
are functionals of the (time-dependent) density. This latter quantity can be
evaluated from the {\em occupied} Kohn-Sham orbitals
\begin{equation}
  \label{eq:tddens}
  n(\vec r, t) = \sum_m^{\rm occ.} |\psi_m(\vec r, t)|^2
  \,.
\end{equation}

We are concerned with external potentials that are the sum of a time-independent
part, typically created by a set of nuclei, and a monochromatic electric field
$v_{\rm field}(\vec r, t)=\sum_{i=1}^3\lambda_i\,r_i\cos\left(\omega{t}\right)$.
If we assume that the magnitude of $\lambda$ is small, we can use perturbation
theory to expand the Kohn-Sham wave-functions in powers of $\lambda$. The first
order term reads
\begin{multline}
%\begin{equation}
  \label{eq:psi}
  \psi_{m}(\vec r, t) = e^{-i\left(\epsilon_m+\sum_{i=1}^3\lambda_i\epsilon^{(1)}_{m\,i}\right){t}}\Big\{
  \psi^{(0)}_m(\vec r) + \\
   \frac12\sum_{i=1}^3\big[\lambda_i e^{i\omega{t}} \psi^{(1)}_{m,\,i}(\vec r, \omega)
     +\lambda_{i}{e^{-i\omega{t}}} \psi^{(1)}_{m,\,i}(\vec r, -\omega)\big]
  \Big\}
  \,,
%\end{equation}
\end{multline}
where $\psi^{(0)}_m(\vec r)$ are the wave functions of the static
Kohn-Sham Hamiltonian $H^{(0)}$ obtained by taking $\lambda=0$
\begin{equation}
  H^{(0)}\psi^{(0)}_m(\vec r) =\epsilon_m\psi^{(0)}_m(\vec r)
  \,,
\end{equation}
and $\psi^{(1)}_{m,\,i}(\vec r, \omega)$ are the first order
variations of the time-dependent Kohn-Sham wave-functions.

From \eqref{eq:psi} and the definition of the time-dependent density
Eq.~\eqref{eq:tddens}, we can obtain the time dependent density

\begin{multline}
%\begin{equation}
  \label{eq:exprho}
  n\dert = n\ord0\der + \frac12\sum_{i=1}^3\big[\lambda_i
  e^{i\omega{t}}n^{(1)}_{i}(\vec r, \omega)\\
  +\lambda_{i}{e^{-i\omega{t}}}n^{(1)}_{i}(\vec r, -\omega)\big]
  \Big\}\,,
%\end{equation}
\end{multline}

\noindent with the definition of the first-order variation of the density
\begin{multline}
%\begin{equation}
  \label{eq:varrho}
  n\ord{1}_i(\vec r, \omega) = \sum_m^{\rm occ.} \Big\{
    \left[\psi\ord0_m\der\right]^*\psi\ord{1}_{m,\,i}(\vec r, \omega)\\
     + \left[\psi\ord{1}_{m,\,i}(\vec r, -\omega)\right]^*\psi\ord0_m\der
  \Big\}\ .
%\end{equation}
\end{multline}

By replacing the expansion of the wave-functions \eqref{eq:psi} in the
time-dependent Kohn-Sham equation \eqref{eq:tdks}, and picking up the first
order terms in $\lambda$, we arrive at a Sternheimer equation for the variations
of the wave functions

\begin{multline}
%\begin{equation}
  \label{eq:sternheimer}
  \left\{H\ord0 - \epsilon_m\pm\omega +
    i\eta\right\}\psi\ord{1}_{m,\,i}(\vec r, \pm\omega) = \\
  -P_cH\ord{1}_i(\pm\omega) \psi_m\ord0(\vec r)
  \,,
%\end{equation}
\end{multline}

with the first order variation of the Kohn-Sham Hamiltonian
\begin{multline}
%\begin{equation}
  H\ord{1}_i(\omega)=
  r_i
  +\mint{r'} \frac{n\ord{1}(\vec{r}',\omega)}{|\vec{r}-\vec{r}'|}\\
  +\mint{r'} f_{\rm xc}(\vec r, \vec r')\,n\ord{1}(\vec{r'}, \omega)
  \,,
%\end{equation}
\end{multline}

\noindent  where \(P_c\) is the projector onto the unoccupied
subspace.  The effect of this projector is to make zero the components
of \(\psi\ord{1}_{m,\,i}(\vec r, \pm\omega)\) in the subspace of the
occupied ground state wavefunctions. In linear response, these
components do not contribute to the variation of the
density\footnote{This is straightforward to prove by expanding the
  variation of the wavefunctions in terms of the ground state
  wavefunctions, using standard perturbation theory, and then
  replacing the resulting expression in the variation of the density
  (Eq.~\ref{eq:varrho}).}, therefore we can safely ignore the
projector.  This is important for large systems as the cost of the
calculation of the projections scales quadratically with the number of
orbitals.

The first term of \(H\ord{1}_i(\omega)\) comes from the external perturbative field,
while the next two represent the variation of the Hartree and
exchange-correlation potentials. The exchange-correlation kernel is a functional
of the ground-state density $n\ord0$, and is given by the functional
derivative

\begin{equation}
  f_{\rm xc}[n\ord0](\vec r, \vec r') = 
  \left.\frac{\delta v_{\rm xc}(\vec r)}{\delta n(\vec r')}\right|_{n = n\ord0}
  \,.
\end{equation}
In the previous equations we made use of the adiabatic approximation to write
$f_{\rm xc}$ as a frequency independent quantity.  Equations \eqref{eq:varrho} and
\eqref{eq:sternheimer} form a set of self consistent
equations for linear response, that only depend on the occupied ground state
orbitals.

Note that we included in Eq.~\eqref{eq:sternheimer} a positive
in\-fi\-ni\-te\-si\-mal $\eta$. This term is essential to obtain the
correct position of the poles of the causal response function, and
therefore to obtain the imaginary part of the
po\-la\-ri\-za\-bi\-li\-ty.  Furthermore, using a small, but finite,
$\eta$ allows us to solve numerically the Sternheimer equation close
to re\-so\-nan\-ces, as it removes the divergences of this equation.

By following the same kind of reasoning, we can arrive at a hierarchy
of Sternheimer equations for the higher order terms in $\lambda$.
These will be needed for the calculation of $\gamma$, the second order
hyperpolarizability, or higher order
hyperpolarizabilities\cite{gonze95}.
  
\subsection{Polarizability}

The time dependent dipole moment is defined as 
\begin{equation}
\label{eq:dipole}
\mu_i(t)=\mint{r} n \left(\vec{r},t\right)r_i\,.
\end{equation}

\noindent The polarizabilities are defined by the expansion of the dipole moment
in terms of the electric field

\begin{multline}
  \label{eq:dipoleexp}
  \mu_i(t)=\mu_i(0)
  +\sum_{j}^3\alpha_{ij}(\omega_j)\lambda_j\cos(\omega_jt)\\
  +\frac1{2!}\sum_{j,k=1}^3\beta_{ijk}(-\omega_j-\omega_k;\omega_j,\omega_k)
  \lambda_j\lambda_k\cos(\omega_jt)\cos(\omega_kt) \\+\ldots
\end{multline}

We must notice that there are several conventions for the definition
of the (hyper)polarizabilities, which are conveniently detailed in
Ref.~\onlinecite{willetts92}. In this work we follow convention AB
(where the prefactors $1/n!$ are explicitly included in
Eq.~\ref{eq:dipoleexp}), that appears to be the most used by the
theoretical community. All referenced values have been converted to
this convention.

If we replace expression (\ref{eq:varrho}) in Eq.~(\ref{eq:dipole})
and compare with (\ref{eq:dipoleexp}) we can obtain a formula for the
polarizability in terms of the variation of the density

\begin{equation}
  \alpha_{ij}(\omega)=\mint{r} n \ord{1}_{j} (\vec r, \omega) r_i
  \,.
\end{equation}

The quantity most easily accessible experimentally is the
photoabsorption cross section, that can be evaluated directly from the
linear polarizability

\begin{equation}
  \bar\sigma(\omega)=
  \frac{4\pi\omega}c\mathfrak{Im}\, \bar\alpha(\omega)
  \,,
\end{equation}

\noindent where \(\bar\alpha\) is the trace of the polarizability tensor

\begin{equation}
  \bar\alpha(\omega)=\frac13\sum_{i=1}^3\alpha_{ii}(\omega)\ .
\end{equation}

\subsection{First hyperpolarizability}
  
If for the dynamic hyperpolarizabilities we follow the same procedure
as before, we get an expression in terms of the second order variation
of the density that requires the evaluation of higher order variations
of the wave functions.  However, it is possible to get the first
hy\-per\-po\-la\-ri\-za\-bi\-li\-ty directly from the first order
variations by means of the $2n+1$ theorem. This theorem states that
the $n$th order variations of the wave functions are enough to obtain
the $2n+1$ derivative of the energy\cite{gonze89,review_baroni}. This
theorem can be expanded to the dynamic case and allows us to write the
first hyperpolarizability $\beta$ in terms of the first order
variations of the wave functions. After some algebra, we arrive
at\cite{rubio96}
\begin{widetext}
\begin{multline}
  \beta_{ijk}(-\w1;\w2,\w3) = -4\sum_P \sum_{\zeta=\pm 1} \Bigg\{
  \sum_m^{occ.}\mint{r} \psi^{(1)*}_{m,\,i}(\vec r, -\zeta\omega_1)
  H\ord{1}_{j}(\zeta\omega_2)
  \psi\ord{1}_{m,\,k}(\vec r, \zeta\omega_3)
  \\
  -\sum_{mn}^{occ.}\mint{r} \psi^{(0)*}_m(\vec r) H\ord{1}_{j}(\zeta\omega_2) \psi^{(0)}_{n}(\vec r)
  \mint{r} \psi^{(1)*}_{n,\,i}(\vec r, -\zeta\omega_1) \psi^{(1)}_{m,\,k} (\vec r, \zeta\omega_3)
  \\
  -\frac{2}{3}\mdint{r}{r'}\!\!\!\mint{r''}
  K\xc(\vec{r},\vec{r}',\vec{r}'')
  n\ord1_{i}(\vec{r},\omega_1) n\ord1_{j}(\vec{r}',\omega_2) n\ord1_{k}(\vec{r}'',\omega_3)\Bigg\}
  \ .
  \label{eq:hpol}
\end{multline}
\end{widetext}
where the first sum is over the permutations $P$ of the pairs
$(i,-\w1)$, $(j,\w2)$, and $(k,\w3)$ and the exchange-correlation
kernel, written in the adiabatic approximation, reads
\begin{equation}
  K\xc(\vec{r},\vec{r}',\vec{r}'') = 
  \left.\frac{\delta^2 v\xc(\vec{r})}{\delta n(\vec{r}')\delta n(\vec{r}'')}
  \right|_{n=n^{(0)}}
  \,.
\end{equation}
      
The first hyperpolarizability tensor has 27 components and is in general
non-symmetric. The quantity that is experimentally relevant is
\begin{equation}
  \bpar = \frac15\sum_{i=1}^{3}\left(\beta_{zii}+\beta_{izi}+\beta_{iiz}\right)
\end{equation}
\noindent Where $z$ is oriented in the direction of the dipole moment of
molecule. Sometimes the equivalent quantity \(\beta_{\rm
  vec}=\beta_{z}=5/3\,\beta_{\parallel}\) is used.

\section{Implementation}
\label{sec:implementation}  
  
This scheme has been implemented using a real space grid-based
formulation in the code {\tt octopus}\cite{octopus03}. We have chosen
a real space grid, as it allows us a systematic convergence of the
results (hyperpolarizabilities are notoriously difficult to converge
with localized basis sets). However, uniform grids can not easily
describe all-electron atoms, so we replace the electron-nuclear
Coulomb interaction by Kleinman-Bylander pseudopotentials. This is,
however, a well controlled approximation for the systems we are
interested in.

In (TD)DFT several approximations exist for the
exchange-correlation term \cite{Scuseria05}. In our approach we can
treat the exchange-correlation term at two levels, one is the ground
state exchange-correlation potential involved in the calculation of
the ground state wave functions, and the other is the
exchange-correlation kernel. For the ground state, except were noted
otherwise, we use the local density approximation (LDA), we will also
use the exact exchange functional in the KLI\cite{KLI}
approximation. For the exchange-correlation kernel, we have decided to
use, due to its simplicity, the adiabatic local density approximation
(LDA). (Although our scheme is quite general and can in principle be
applied to any exchange-correlation functional.)

The LDA is a well studied approximation, and is quite reliable
in the prediction of many properties. One important main defect in
this context is the wrong asymptotic part of the LDA
exchange-correlation potential, that for neutral systems decays
exponentially instead of falling as $1/r$. This usually leads to small
HOMO-LUMO gaps which implies systems that are too polarizable, in
contrast to Hartree-Fock where the gap is larger and the magnitudes of
the polarizabilities are underestimated. The exchange and correlation
kernel will contribute to reduce the independent particle
polarizability even at ALDA level (this contribution for extended
systems is zero and as the long range behavior of the XC potential is
not relevant in this regime, this clearly points that the non-locality
of the XC kernel as well as self-interaction correction are
responsible for the bad performance of LDA). We will observe this
overestimation in the calculations that follow. In fact, in the case
of compact finite systems there has been indications that the
exchange-correlation potential seems to be more important than the
kernel\cite{Stener2001}. The situation is particularly problematic in
the case of long molecular chains\cite{5.4.vanGisbergen}, where
standard exchange-correlation func\-tio\-nals can greatly overestimate
polarizabilities when compared to many-body approaches. Note, however,
that this is not a deficiency of DFT, but of the LDA approximation
(and of many exchange-correlation functionals), that can in principle
be treated\cite{Casida00} by using more sophisticated orbital
dependent functionals like the self-interaction corrected LDA
\cite{ldapz} or the exact exchange\cite{goerling94}. However in
present orbital functionals there is still a significant correlation
contribution that is not taken into account and is responsible for the
discrepancies between theory and experiment for long-chains in the
exact exchange approach\cite{5.4.vanGisbergen}.

Numerically, the central part of our scheme is the solution of the
Sternheimer equation \eqref{eq:sternheimer}. This has the form of a
linear equation were the operator to invert is the shifted ground
state Hamiltonian. As the shift is complex (due to the $i\eta$ term),
this operator is not Hermitian. Therefore, we can not use standard
techniques common in the community, like the simple conjugated
gradients scheme, but have to rely on more general (and involved)
linear solvers. Our choice was the {\it biconjugate gradient
  stabilized} method\cite{Saad}. Close to the resonance frequencies,
the Sternheimer equation becomes very badly conditioned, and the
solution process turns out to be very costly. The problem can be eased
by the use of preconditioning.  We have found that a smoothing
preconditioner\cite{saad96} can dramatically improve convergence in
these cases. Also for small system the solution process can be made
less costly if we solve the Sternheimer equation in the space of the
unoccupied wave functions by orthogonalizing the right-hand side of
the Sternheimer equation with respect to the occupied wave functions.
Although this would not be practical for large systems as the
orthogonalization process can become very demanding.Even with
these techniques, the process is much more costly for frequencies near
resonance. For example, for the CO case, the full self-consistent
solution of the Sternheimer equation for a single frequency in a
single direction requires around \(1700\) applications of the
Hamiltonian, for the a near resonance frequency approximately 10 times
more applications are required. As a comparison, for the ground state
DFT calculation around \(2000\) Hamiltonian operations are needed.

As the right-hand side of the Sternheimer equation depends on the
linear variation of the density, the problem has to be solved
self-consistently. For this, we use similar strategies as for the
ground-state calculation, mixing the linear variation of the density
using a Broyden scheme\cite{Broyden65} in order to speed up the
convergence of the self-consistent cycle.

The Poisson equation is solved using the interpolating scaling
functions scheme proposed in Ref.~\onlinecite{isf}. As this process
has to be done only once per self-consistency iteration, the
performance of the Poisson solver is not critical.

The total cost of calculating the response is of order ${\cal
  O}(N_{ks}N_gM_{\omega})$, where $N_{ks}$ is the number of Kohn-Sham
orbitals, $N_g$ is the number of grid points and $M_\omega$ is the
number of frequencies we desire (which is independent of the system
size). This scaling is much better than for the approaches that relay
on expansions in particle-hole states, and even better than for the
ground-state calculation that normally scales as ${\cal
  O}(N_{ks}N^2_g)$ (due to the necessity of orthogonalizing the
wave-functions). We believe, therefore, that this method can be used
to study (hyper)polarizabilities of very large systems, like
nanostructures or molecules with biological interest. After obtaining
the linear response, the evaluation of the hyperpolarizability from
Eq.~(\ref{eq:hpol}) has a cost proportional to ${\cal
  O}(N^2_{ks}N_gM_{\omega})$ but with a very small prefactor. Note that
\({\cal O}(N)\) (with $N$ the number of atoms) schemes are available
for ground state\cite{ordern} and static polarizability
calculations\cite{ordernpol}. These linear scaling methods are based
on ``nearsightness''\cite{kohn} or the idea of divide and
conquer\cite{yang}. The real space treatment used in our method would
allow for the incorporation of these strategies to reduce even further
the numerical cost for large systems.

\section{Examples of applications}
\label{sec:applications}

We decided to illustrate the implementation of our method by applying
it to some simple molecules that have been well studied both
experimentally and theo\-re\-ti\-ca\-lly: CO, H$_2$O, and
paranitroaniline (PNA).

In all our calculations we took the experimental geo\-me\-tries:
$\overline{\rm CO}=1.13$\,\AA; $\overline{\rm HO}=0.957$\,\AA, ${\rm
  \hat {HOH}}=104.5^{\rm o}$; the experimental geometry of PNA
determined through X-ray crystallography can be found in
Ref.~\onlinecite{bertinelli77}.  However, there are two CH distances
missing from the crystallographic data. For these we followed
Ref.~\onlinecite{salek05} and took their theoretical values (see
Table~1 of Ref.~\onlinecite{karna91}). In all cases the dipole moment
of the molecule was taken perpendicular to the \(z\) axis, for H$_2$O
the molecule was considered in the $yz$ plane and in the case of PNA
it was taken in the $xz$ plane. We used Troullier-Martins
norm-conserving pseudopotentials with a core radius of 0.66 \AA{} for
H, 0.78 \AA{} for C and 0.74 \AA{} for both N and O.

We used a simulation box composed of spheres centered at each atomic
position, with a radius of 9.5\,\AA\ for CO, 7.4\,\AA\ H$_2$O and
5.3\,\AA\ for PNA. The points were distributed in a regular grid with
spacing 0.20\,\AA\ for CO, 0.17\,\AA\ for H$_2$O, and 0.19\,\AA\ for
PNA. With these pa\-ra\-me\-ters, hyperpolarizabilities are converged
to better than 1\%. As expected, the simulation boxes required to
converge the hyperpolarizabilities were much larger than the ones
typically used in ground-state calculations, as these quantities have
sizeable contributions from the regions far away from the nuclei. The
required simulation box size depends on the frequency of the
perturbation, being larger for higher frequencies. Finally, we used
the LDA parameterization by Perdew and Zunger\cite{ldapz} to
approximate the exchange-correlation functionals.

\begin{table}
  \centering
  \begin{tabular}{lrrrrrr}
                 & This work    &  LDA & HF$^a$ & MP4$^a$ & CCSD(T)$^a$ &	Exp.	\\
    \hline \\[-3mm]
    $\mu	$ &	0.0631	&&	-0.1052	& 0.0905&	0.057	&	0.0481$^b$	\\
    $\alpha_{xx}$ &	12.55	&&	11.25	& 12.00	&	11.97	&		\\
    $\alpha_{zz}$ &	15.82	&&	14.42	& 15.53	&	15.63	&		\\
    $\bar\alpha	$ &	13.64	& 13.87$^c$ & 12.31	& 13.18	&	13.19	&	13.09$^d$	\\
    $\beta_{xxz}$ &	8.35	& 8.24$^e$ 	&	5.0 & 8.3	&	8.4	&		\\
    $\beta_{zzz}$ &	33.34	& 33.52$^e$ &    31.1 & 28.3	&	30.0	&		\\
    $\bpar	$ &	30.03	& 30.00$^e$  &	24.8	&	 27.0	&	28.0	&		\\
  \end{tabular}
  \caption{Comparison of static polarizabilities and
    hyperpolarizabilities for CO. Results are in atomic units.
    $^a$Finite differences results from Ref.~\onlinecite{hp_co}.
    $^b$Experimental result from Ref.~\onlinecite{mu_co}.
    $^c$LDA basis set results from Ref.~\onlinecite{vangisbergen98a}.
    $^d$Experimental result from Ref.~\onlinecite{pol_co}.
    $^e$LDA basis set results from Ref.~\onlinecite{vangisbergen98b}.
  }
  \label{tab:static_co}
\end{table}

We will start our discussion with CO. The results for static properties are
displayed in Table~\ref{tab:static_co}. The results fully agree with the other
DFT-LDA calculations. Compared to more sophisticated quantum chemistry methods,
the LDA overestimates the values for the polarizabilities and the
hyperpolarizabilities.  This, as mentioned before, comes from a deficiency in
the asymptotic region of the LDA potential.

\begin{figure}[t]
  \centering
  \includegraphics*[width=9cm]{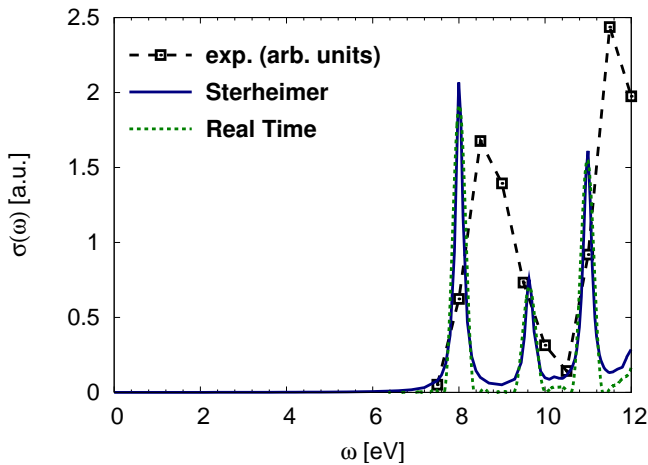}
  \caption{Average photo-absorption cross section of the CO molecule, calculated
    within the adiabatic LDA. The line corresponds to the results obtained
    through the solution of the Sternheimer equation, while the dots are
    obtained through the solution in real-time of the time-dependent Kohn-Sham
    equation. Low resolution experimental absorption cross-section from
    Ref.~\onlinecite{Chan93}. For a more detailed comparison, the relevant
    experimental excitation energies are at: 8.51\,eV ($A^1\Pi$), 10.78\,eV
    ($B^1\Sigma^+$), 11.40\,eV ($C^1\Sigma^+$), and 11.53\,eV
    ($E^1\Pi$)\cite{Chan93,Nielson80}.  }
  \label{fig:pol_co}
\end{figure}

We now turn to the dynamic properties. In Fig.~\ref{fig:pol_co} we
plot the absorption cross section of the CO molecule obtained with our
approach, and compared to the spectrum obtained through the direct
solution of the time-dependent Kohn-Sham equations in real-time. As
expected, the two calculations agree perfectly, which validates our
numerical implementation of the Sternheimer equations. Note that, even
if both methods have identical scalings, the solution of the
Sternheimer equation still has a larger prefactor if the whole
spectrum is required. This is due to the ill-conditioning of the
linear system close to resonances. From Fig.~\ref{fig:pol_co} it is
clear that the two theoretical results are red-shifted with respect to
the experimental curve. This shortcoming of the simple LDA can be
corrected by using functionals with the correct asymptotic behavior
\cite{Casida00}.

\begin{figure}[t]
  \centering
  \includegraphics*[width=9cm]{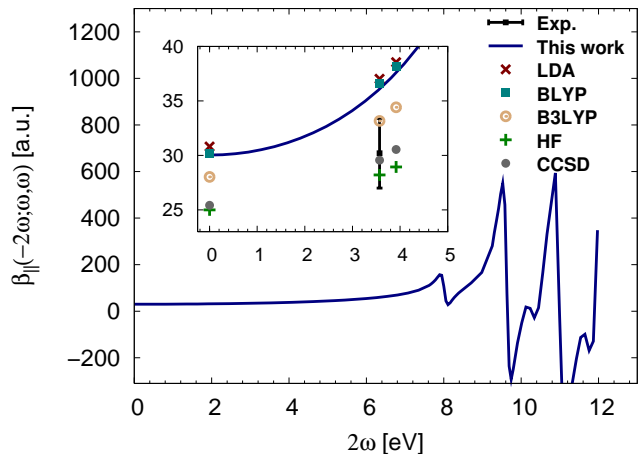}
  
  \caption{(Color online) Second harmonic generation
    $\beta_{||}(-2\omega;\omega,\omega)$ of CO.  The inset shows a
    comparison of the results of this work (TDLDA) with other
    available results.  Exp: Experimental results from
    Ref.~\onlinecite{shelton94}; HF: Hartree Fock results from
    Ref.~\onlinecite{salek05}; LDA, BLYP, B3LYP: DFT results from
    Ref.~\onlinecite{salek05}; CCSD: Coupled cluster results from
    Ref.~\onlinecite{salek05}.  }\label{fig:hp_dyn_co}
\end{figure}

In Fig.~\ref{fig:hp_dyn_co} we present our calculations of the second
harmonic generation spectrum, $\beta_{||}(-2\omega;\omega,\omega)$, of
CO, together with the available experimental results\cite{shelton94}
and previous theoretical data\cite{salek05}. Our results agree very
well with previous DFT results using the LDA. We can also see that the
use of the generalized gradient approximation BLYP (Becke 88
\cite{Becke88} for exchange and Lee, Yang, and Parr \cite{LYP88}
correlation) does not change significantly the results. \-Using a hybrid
functional, the Becke 3 parameter B3LYP functional \cite{Becke93},
does reduce the error, while Hartree-Fock (HF) results underestimate
the value for the hyperpolarizability. The best results are, as
expected, obtained by coupled cluster calculations using singles and
doubles (CCSD).
\begin{table}[t]
  \centering
  \begin{tabular}{lrrrrrr}
    & \multicolumn{2}{c}{\bf $\bm\omega=$0.00\,eV} & \multicolumn{2}{c}{\bf
      $\bm\omega=$1.79\,eV} &
    \multicolumn{2}{c}{\bf $\bm\omega=$1.96\,eV} \\
    & This work  &  LDA$^a$	& This work  &  LDA$^a$	& This work  &  LDA$^a$
    \\
    \hline \\[-3mm]
    $\beta_{zzz}$ & -21.23 & -19.14 & -27.67 & -25.22 & -29.37 & -26.81 \\
    $\beta_{zxx}$ &  -9.84 &  -8.82 & -12.89 & -11.45 & -13.68 & -12.11 \\
    $\beta_{xxz}$ &  -9.84 &  -8.82 & -16.87 & -15.57 & -19.28 & -17.90 \\
    $\beta_{zyy}$ & -12.08 & -11.67 & -14.94 & -14.39 & -15.68 & -15.09 \\
    $\beta_{yyz}$ & -12.08 & -11.67 & -14.48 & -13.99 & -15.06 & -14.56 
  \end{tabular}
  \caption{Tensor components for second harmonic generation $\beta(-2\omega;\omega,\omega)$
    of H$_2$O. $^a$Theoretical results from Ref.~\onlinecite{salek05}.
  }\label{tab:shg_h2o}
\end{table}

Now we turn to the H$_2$O molecule. To test our implementation, we show, in
Table~\ref{tab:shg_h2o}, the different components of the hyperpolarizability
tensor for $\omega=0$\,eV, 1.79\,eV, and 1.96\,eV. We see that our results
compare well to previous theoretical work~\cite{salek05}.  The small difference
can be explained by the different numerical methodologies (real-space grid and
pseudopotentials in our case and basis sets in Ref.~\onlinecite{salek05}).

\begin{table*}
  \centering
  \begin{tabular}{lrrrr}
    &\(\alpha(0,0)\) & 
    \(\beta_{\parallel}(0;0,0)\) & 
    \(\beta_{\parallel}(0;\omega,-\omega)\) &
    \(\beta_{\parallel}(-2\omega;\omega,\omega)\) \\
    \hline \\[-3mm]
    This work (TDLDA) & 10.51 & -25.89 & -28.33 & -34.71 \\
    This work (KLI/ALDA) & 8.61 & -11.75 & -12.43 & -14.03 \\
    LDA$^a$ & 10.5 & -26.1 & -28.6 & -35.1 \\
    BLYP$^a$ & 10.8 &-27.9 & -30.9 & -38.8 \\
    LB94$^a$ & 9.64 &-17.8 & -17.7 & -20.3 \\ 
    LDA$^b$	 & 10.63 & -23.78 & -26.09 & -32.12 \\
    BLYP$^b$	& 10.77	& -23.65 & -26.11 & -32.76 \\
    B3LYP$^b$ & 9.81 & -18.54 & -20.11 & -24.11 \\ 
    HF$^b$  & 8.53 &-10.73 & -11.27& -12.52 \\ 
    CCSD(T)$^c$ & 9.79 & -18.0	& -19.0 & -21.1 \\
    exp. & 9.81$^d$ & & & -22\(\pm\)9$^e$\\
  \end{tabular}
  \caption{Comparison of (hyper)polarizabilities of
    H$_2$O for different DFT calculations. CCSD(T) and
    experimental values are given as reference. For dynamic results
    \(\omega=1.79\ [\mbox{eV}]\).
    $^a$Grid based calculations from Ref.~\onlinecite{bertsch01}; 
    $^b$Basis set calculations from Ref.~\onlinecite{salek05};
    $^c$Results from Ref.~\onlinecite{sekino93}; 
    $^d$Experimental results from Ref.~\onlinecite{spackman89}; 
    $^e$Experimental results from Ref.~\onlinecite{ward79}; 
  }\label{tab:h2o_comp}
\end{table*}

In Table~\ref{tab:h2o_comp} we show the static polarizability,
$\alpha(0,0)$, the static first hyperpolarizability,
$\beta_{\parallel}(0;0,0)$, the optical rectification,
$\beta_{\parallel}(0;\omega,-\omega)$, and second harmonic generation,
$\beta_{\parallel}(-2\omega;\omega,\omega)$ for water. All dynamic
values were calculated for $\omega=1.79$\,eV. This time we also
have included results with the KLI orbital dependent
exchange-correlation potential combined with the ALDA
exchange-correlation kernel. Concerning LDA values, we can see that
our results fully agree with the results of
Ref.~\onlinecite{bertsch01} that also uses a grid based
representation, while basis-set results from Ref.~\onlinecite{salek05}
differ by less than 10\%. We can see that all LDA and BLYP values
overestimate the magnitude of (hyper)polarizabilities with respect to
experiment and coupled cluster calculations. The use of more
sophisticated exchange correlation functionals, like LB94 or B3LYP,
improves significantly the results, while the KLI/ALDA scheme gives an
underestimation of the magnitude of (hyper)polarizabilities similar to
Hartee-Fock results.

\begin{figure}[t]
  \centering
  \includegraphics*[width=9cm]{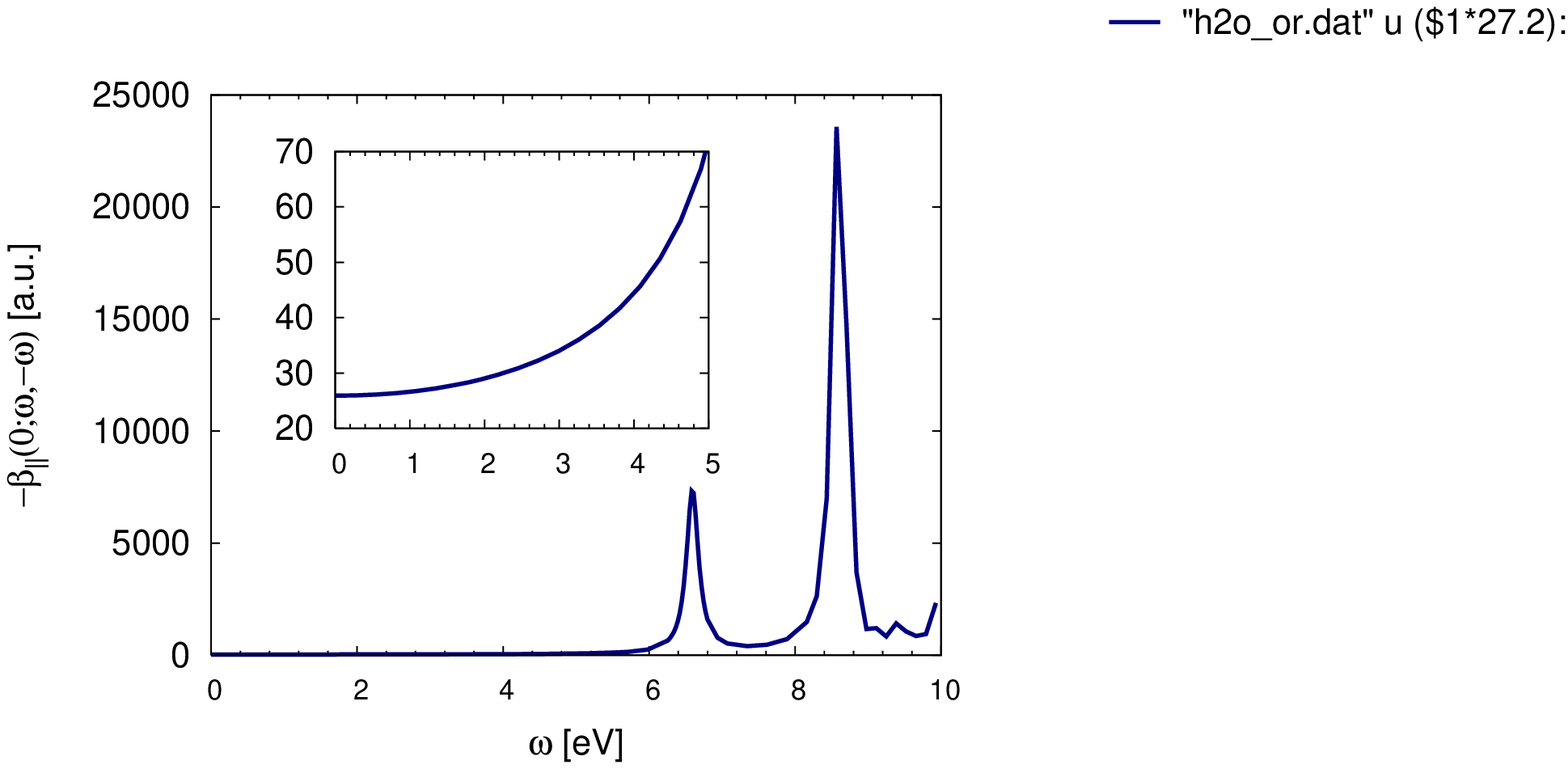}
  \caption{Calculated optical rectification
    \(\beta_{\parallel}\left(0;\omega;-\omega\right)\) of H\(_2\)O.}
  \label{fig:h2o_or}
\end{figure}

To conclude the discussion of water results, we plot, in
Fig.~\ref{fig:h2o_or} the frequency dependence of the optical
rectification in the visible and near ultraviolet regimes. It is clear
that our method not only works for small non-resonant frequencies but
also in the more complicated resonant regime.

\begin{figure*}[t]
  \centering
  \includegraphics*[width=8.1cm]{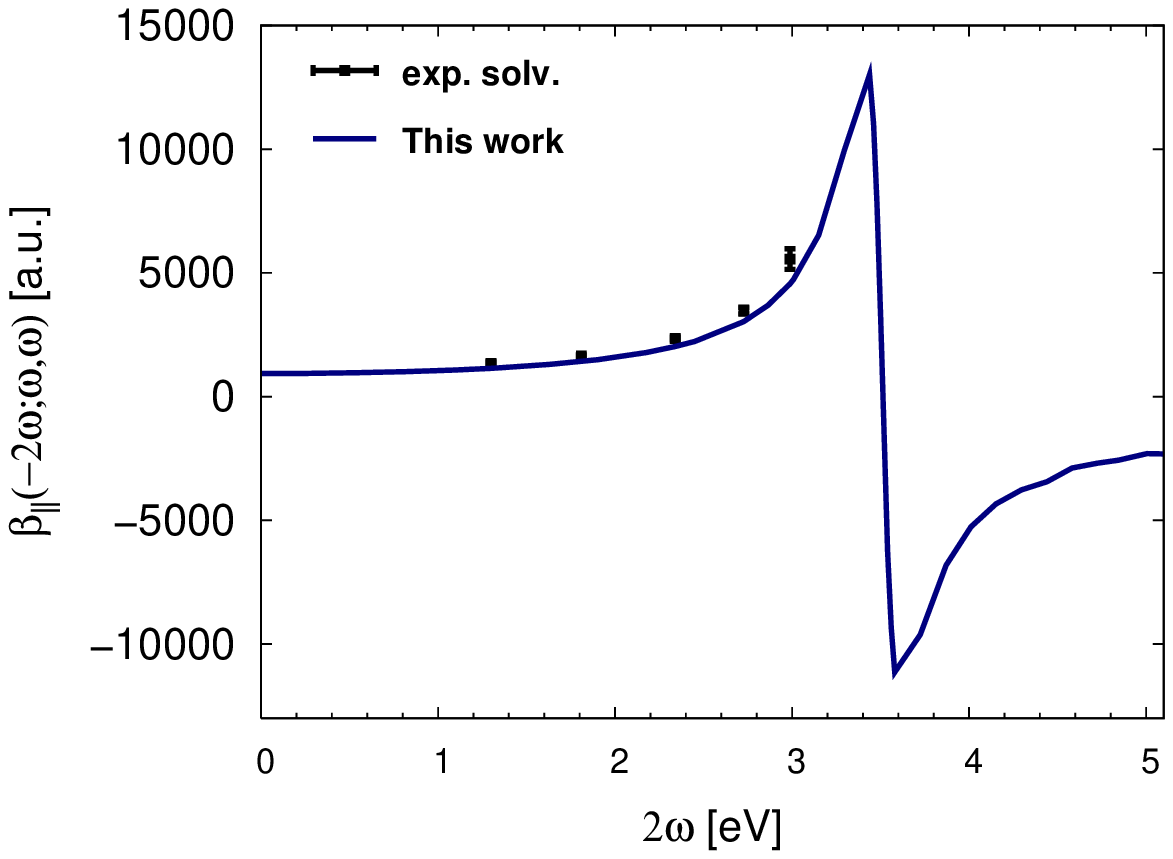}
  \includegraphics*[width=8.1cm]{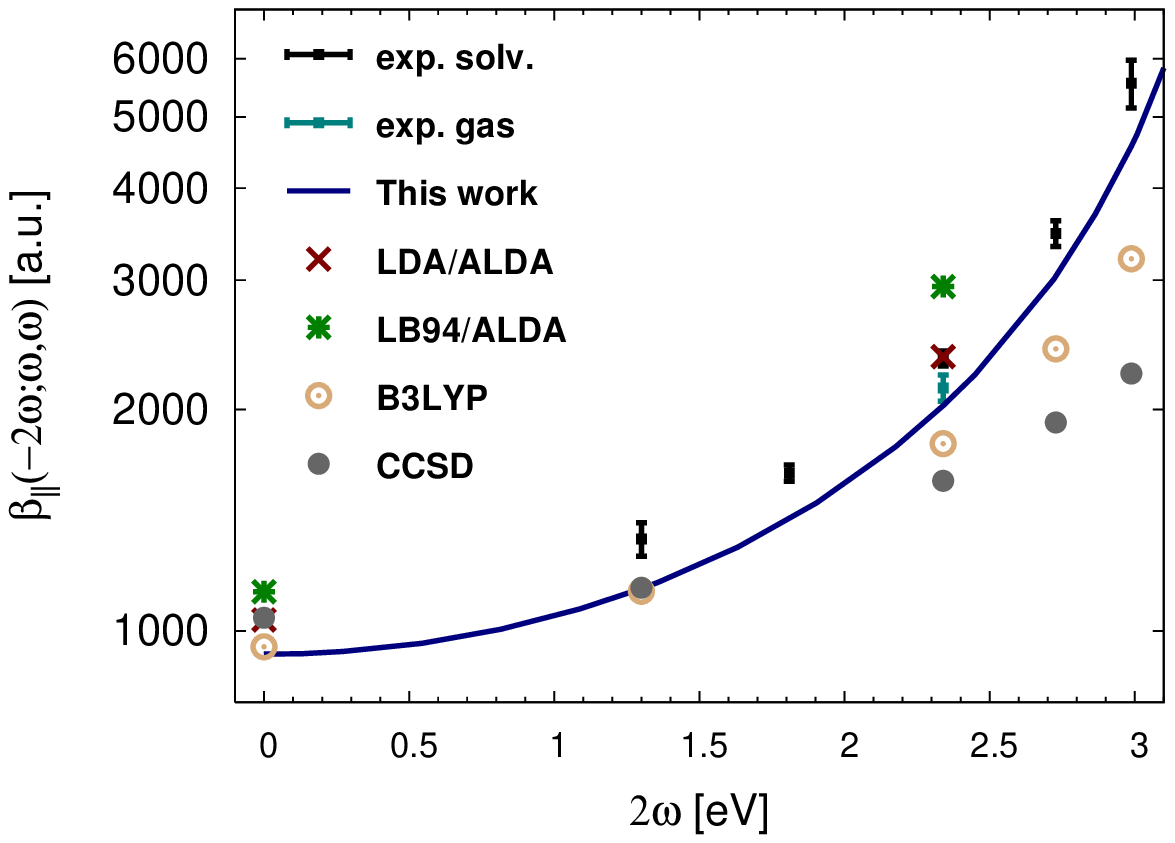}
  \caption{(Color online) Second harmonic generation
    $\beta_{||}(-2\omega;\omega,\omega)$ of paranitroaniline. Note
    that the $y$ axis in the right panel is in logarithmic scale.
    exp. solv.: Solvent phase experimental results from
    Ref.~\onlinecite{teng83}; exp. gas: Gas phase experimental results
    from Ref.~\onlinecite{kaatz98}; LDA/ALDA, LB94/ALDA: DFT basis set
    results from Ref.~\onlinecite{vangisbergen98b}; B3LYP: DFT basis
    set results from Ref.~\onlinecite{salek02}; CCSD: Coupled cluster
    results from Ref.~\onlinecite{salek05}. Some references use a
    different convention to define hyperpolarizabilities, all the
    values shown here have been converted to convention
    AB.}\label{fig:pna_shg}
\end{figure*}

Finally, we turn to a larger molecule: paranitroaniline. Our results
for the second harmonic generation process in this molecule are given
in Fig.~\ref{fig:pna_shg}. There are two ex\-pe\-ri\-ments
a\-vai\-la\-ble: i)~a gas phase experiment \cite{kaatz98} performed
for a single frequency; and ii)~paranitroaniline in solvent for
several frequencies\cite{teng83}.  The latter were corrected for the
presence of the solvent, but this correction is clearly incomplete as
the value from Ref.~\onlinecite{teng83} for $\omega=1.17$\,eV is still
$10\%$ larger that the gas phase measurement\cite{kaatz98}. We include
also other theoretical values using DFT\cite{vangisbergen98b,salek02}
and CCSD\cite{salek05}. We can observe that our results underestimate
the solvent ex\-pe\-ri\-men\-tal results by about 15\% for all
available frequencies. In comparison, B3LYP and CCSD results are
seriously too small at high frequencies, with values that are around
40-50\% smaller than experiment. We think that the reason for this
discrepancy is the better description of the
hy\-per\-po\-la\-ri\-za\-bi\-li\-ties near resonance of our method.
Furthermore, it uses a grid based representation that describes better
the regions far from the nuclei in comparison with localized basis
sets, allowing for larger flexibility in capturing the dynamic changes
in the wave functions.

\section{Conclusions}
\label{sec:conclusions}
  
In summary, we have presented a method that allows the calculation of
both static and dynamic polarizabilities and hyperpolarizabilities.
Our approach is based on the Sterheimer equation, within the formalism
of time-dependent density functional theory, and requires the solution
of an non-Hermitian linear equation. This solution is obtained through
a ge\-ne\-ra\-li\-za\-tion of the conjugated gradients method using a
real-space (basis set free) representation of the wave-functions. In
this way we are able not only to obtain static quantities, but also
the whole frequency dependence of the (hyper)polarizabilities even
close to resonances.  The scaling with the number of atoms in the
system is excellent, so we expect that the method will be useful for
the study of very large systems. First applications to small benchmark
molecules yield quite good results in comparison to previous
theoretical approaches and experimental results.

One of the beauties of this approach is how easily it can be
generalized to higher orders and to handle other kinds of static or
dynamic perturbations.  For example, phonon frequencies, (resonant)
Raman tensors, NMR tensors, forces in the excited state, etc. can all
be obtained by just changing the right-hand side of the Sternheimer
equations. The third and higher order polarizabilities can also be
obtained by solving a hierarchy of Sternheimer equations that have the
same form as the first order one. Work has already started to extend
our implementation in these directions.

\begin{acknowledgments}
  The authors were partially supported by the EC Network of Excellence
  NANOQUANTA (NMP4-CT-2004-500198), SANES project
  (NMP4-CT-2006-017310), DNA-NANODEVICES (IST-2006-029192), NANO-ERA
  Chemistry, MCyT and Barcelona Supercomputing Center (Mare Nostrum).
  XA would also like to acknowledge partial support from the EU
  Programme Marie Curie Host Fellowship (HPMT-CT-2001-00368).
\end{acknowledgments}

%\bibliography{hp}

\end{document}